\documentclass[sigconf,10pt]{acmart}
\setcopyright{acmlicensed}
\copyrightyear{2026}
\acmYear{2026}
\acmDOI{XXXXXXX.XXXXXXX}
\acmConference[Submitted to MobiSys '26]{}{June 21--25,
  2026}{Cambridge, UK}

\acmISBN{978-1-4503-XXXX-X/2026/07}

\renewcommand\footnotetextcopyrightpermission[1]{} 

\newcommand{\name}{FedWiLoc\xspace}

\usepackage[english]{babel}
\usepackage{blindtext}
\usepackage{xcolor}
\usepackage{amsfonts}
\PassOptionsToPackage{hyphens}{url}\usepackage{hyperref}
\usepackage{graphicx}
\usepackage{url}
\usepackage{listings}
\usepackage{multicol}
\usepackage{multirow}
\usepackage{rotating}
\usepackage{xspace}
\usepackage[ruled,vlined]{algorithm2e}
\usepackage{comment}
\usepackage{enumitem}
\usepackage{amsmath}
\usepackage{mathrsfs}
\usepackage{cancel}
\usepackage{cleveref}
\usepackage{subcaption}

\begin{document}
\title{\name: Federated Learning for Privacy-Preserving WiFi Indoor Localization}

\author{Kanishka Roy}
\authornote{Both authors contributed equally to this research.}
\affiliation{%
  \institution{UC Los Angeles}
  \state{California}
  \country{USA}
}
\email{kroy02@ucla.edu}

\author{Tahsin Fuad Hasan}
\authornotemark[1]
\affiliation{%
  \institution{University at Buffalo-SUNY}
  \state{New York}
  \country{USA}
}
\email{tahsinfu@buffalo.edu}

\author{Chenfeng Wu}
\affiliation{%
  \institution{Latitude AI, Detroit}
  \state{Michigan}
  \country{USA}
}
\email{cwu@lat.ai}

\author{Eshwar Vangala}
\affiliation{%
  \institution{University at Buffalo-SUNY}
  \state{New York}
  \country{USA}
}
\email{eshwarsa@buffalo.edu}

\author{Roshan Ayyalasomayajula}
\affiliation{%
  \institution{University at Buffalo-SUNY}
  \state{New York}
  \country{USA}
}
\email{roshana@buffalo.edu}

\renewcommand{\shortauthors}{Roy et al.}

\begin{abstract}
Current data-driven Wi-Fi-based indoor localization systems face three critical challenges: protecting user privacy, achieving accurate predictions in dynamic multipath environments, and generalizing across different deployments.
Traditional Wi-Fi localization systems often compromise user privacy, particularly when facing compromised access points (APs) or man-in-the-middle attacks. As IoT devices proliferate in indoor environments, developing solutions that deliver accurate localization while robustly protecting privacy has become imperative.
We introduce \name, a privacy-preserving indoor localization system that addresses these challenges through three key innovations. First, \name employs a split architecture where APs process Channel State Information (CSI) locally and transmit only privacy-preserving embedding vectors to user devices, preventing raw CSI exposure. Second, during training, \name uses federated learning to collaboratively train the model across APs without centralizing sensitive user data. Third, we introduce a geometric loss function that jointly optimizes angle-of-arrival predictions and location estimates, enforcing geometric consistency to improve accuracy in challenging multipath conditions.
Extensive evaluation across six diverse indoor environments spanning over 2,000 sq. ft. demonstrates that \name outperforms state-of-the-art methods by up to 61.9\% in median localization error while maintaining strong privacy guarantees throughout both training and inference. 
\end{abstract}




\maketitle
\section{Introduction}\label{sec:intro}

Indoor localization and navigation has grown in popularity due to the increase in deployment of Industrial IoT, medical devices, and indoor robots~\cite{iot_trend}. Specifically, indoor localization using wireless signals~\cite{bloc,zhao2021uloc,yang2022vuloc,kotaru2015spotfi,picoscenes}, particularly Wi-Fi Channel State Information (CSI)~\cite{kotaru2015spotfi,picoscenes}, has emerged as a promising approach to enable context-aware applications in robotics, human activity detection, and navigation~\cite{arun2024wais,WiVi,zhang2021widar3,arun2022p2slam,dloc,zhang2024rloc}. It is expected to have a market share of \$43.2 billion by 2030~\cite{locbiz}.

\begin{figure}
    \centering
    \includegraphics[width=\linewidth]{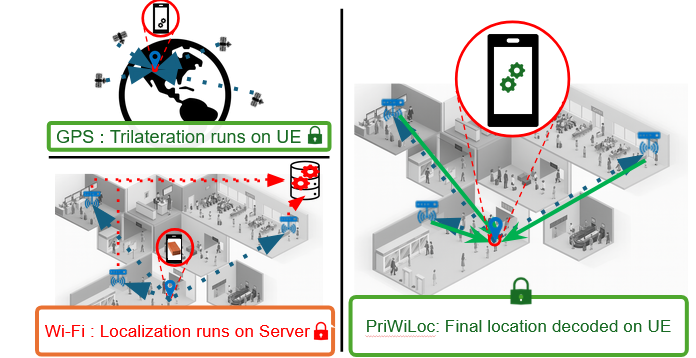}
    \caption{Current systems send the user's measurement across different Wi-Fi Access Points(APs) to a central server to computes the user location. In contrast in \name has a Federated learning architecture that enables the user device to decode its location like GPS.}
    \label{fig:intro_fig}
\end{figure}

Enabling this increase in popularity has been the growth in open-sourced toolboxes~\cite{xie2018swan,picoscenes,arun2024wais,pizarro2021accurate,zhang2024rloc} from academia and changes in Wi-Fi standards~\cite{du2024overview,au2016latest,picazo2023ieee}, making Wi-Fi based indoor localization more accessible. Furthermore, Wi-Fi localization problems have evolved from RSSI (received signal strength) based~\cite{radar,zee,ez} to CSI-based solutions~\cite{chronos,kotaru2015spotfi}, creating effective systems to combat Non-Line of Sight (NLoS) and multipath scenarios. This has led to state-of-the-art deep learning and data-driven solutions~\cite{dloc,zhang2024rloc} which can accurately predict locations in the midst of strong multipath, reflections, and obstructions.

Current learning-based Wi-Fi localization systems follow a centralized architecture (Fig.~\ref{fig:intro_fig}): user devices' wireless channels are measured at multiple APs, transmitted to a central server for location estimation, and the computed location is returned to the user. While this approach simplifies deployment and avoids edge device computation constraints, it introduces critical vulnerabilities. First, users lose control over their location data, creating privacy risks. Centralized location databases can be exploited by adversaries~\cite{wipeep, li2024rftrack} to infer behavioral patterns, conduct surveillance, or enable targeted attacks. Even well-intentioned service providers may be compelled to share location data with third parties or fall victim to data breaches. Second, centralized architectures hinder generalization. When all CSI data is processed at a central server trained on specific environments, deploying to new spaces requires extensive retraining and site-specific parameter tuning. This environment-specific nature makes scalable deployment across diverse indoor spaces impractical, limiting real-world adoption of learning-based localization systems.

To address these challenges, we propose \name, a novel framework that integrates \textbf{federated learning} and environment-agnostic \textbf{geometric losses} into wireless localization.
Our approach enables users to collect embeddings from multiple Wi-Fi Access Points (APs) and decode their own location locally, without transmitting data to third-party servers. This preserves user privacy while leveraging the diversity of wireless environments across deployment sites. Unlike prior work, \name is designed to be \textit{privacy-preserving and robust to environmental variations}, representing a significant step toward practical real-world deployment.

To enforce location privacy, we draw inspiration from GPS: GPS devices perform trilateration using signals from satellites at known positions, computing location entirely on-device. Similarly, \name employs a split encoder-decoder architecture where APs run encoders locally and transmit only embedding vectors to user devices, which then perform decoding. However, simply splitting the model is insufficient—naive approaches are vulnerable to membership inference attacks that can leak device information~\cite{pruthi2020estimating,hu2022membership}.

We address this vulnerability through an inverted federated learning approach inspired by FedAvg~\cite{fedavg}. Rather than a central server aggregating data from multiple users (traditional federated learning), in \name each user aggregates encoded data from multiple APs. The key insight is that by federating encoder training across APs using gradient averaging, we prevent AP-specific encoder fingerprints that could enable user tracking. Each user maintains a personal decoder that, combined with known AP locations, computes their position locally. This architecture enforces privacy during both training and inference by ensuring no raw CSI leaves the APs and no location data is centralized. The natural next question is: how to make the architecture work across diverse environments? The goal is to provide private localization and also robust estimates that can generalize across different environments, router configurations, and bandwidths.

Indoor environments present significant heterogeneity: routers vary in configuration and bandwidth, spaces differ in layout and furniture arrangement, and multipath propagation effects change with human movement. For practical deployment, \name must generalize across these variations without environment-specific retraining. Unlike DLoc~\cite{dloc}, which uses likelihood profiles in Cartesian coordinates tied to specific room dimensions, we represent CSI as 2D Angle-of-Arrival (AoA) versus relative Time-of-Flight (rToF) images. This representation is fundamentally location-agnostic—AoA-ToF images depend only on the signal properties, not the AP's absolute position or the environment's size. This enables \name to scale to arbitrarily large spaces without architectural modifications. To handle varying bandwidths (20 MHz to 160 MHz), we apply subsampling and time-offset slope compensation along the ToF axis, maintaining consistent pixel resolution across all bandwidth configurations. To enforce generalization, we introduce a geometric loss function that jointly optimizes angle predictions and location estimates, enforcing geometric consistency across all APs. This constraint prevents the model from overfitting to environment-specific patterns by requiring predictions that satisfy fundamental geometric relationship between multiple AoAs from different APs. Together, these design choices enable \name to generalize across environment sizes, AP configurations, and bandwidth settings without fine-tuning. 

Finally, we tested \name on the most widely used and open-sourced localization datasets provided by the authors of DLoc~\cite{dloc} and RLoc~\cite{zhang2024rloc}. Across both of these datasets we have trained and tested on 6 environments, 6 user devices, 10 different access point locations, and at both 80~MHz and 40~MHz bandwidths. We compare \name's results with both the state-of-the-art traditional algorithm, SpotFi~\cite{kotaru2015spotfi} and data-driven algorithm, DLoc~\cite{dloc}. 
Across all of these evaluations we have demonstrated that \name achieves the following localization results:

\begin{itemize}[leftmargin=*]
    \item When trained on three scenarios and tested on an unseen scenario, \name achieves 63 cm median and 213 cm (90th percentile) localization error, \textbf{outperforming SpotFi by 59\% and DLoc by 51\% in median error}.
    \item \name demonstrates strong cross-environment generalization: when trained on any 5 environments and tested on the held-out 6th environment, median localization error ranges from 0.8m to 2m across all six train/test combinations.
    \item \name generalizes across all six environments: when trained on portions of all environments and tested on unseen areas, \name achieves a \textbf{median localization error of 55 cm and 90th percentile error of 155 cm}.
\end{itemize}

\section{Motivating Private, AoA-Based Localization}\label{sec:motivation}

In this section, we motivate the necessity for private, AoA-based WiFi localization.

\subsection{Motivation for Private Localization}

Traditional WiFi localization architectures create a fundamental privacy problem: to determine their position, users must reveal their location to the network infrastructure. For GPS, this is not a concern since GPS signals are decoded locally on user devices without the need of processing on a central server. In Wi-Fi, the centralized approach of sending a location to the network enables the continuous tracking of individuals within a building without explicit consent. The location traces collected can reveal sensitive information including daily routines, social interactions, and behavioral patterns -- data that could be exploited for surveillance, targeted advertising, or discriminatory practices. 

Consider the airport scenario in Figure \ref{fig:intro_fig}, where thousands of passengers rely on indoor navigation, service robots shuttle travelers to gates, and autonomous security drones patrol terminals. All of these devices routinely exchange wireless signals with the building’s Wi-Fi infrastructure. In today’s Wi-Fi-based and learning-driven localization systems, user devices transmit location-revealing measurements to access points so that the network can compute their positions. This creates a fundamental vulnerability: any adversary with visibility into the network — via a compromised access point, passive packet sniffing, rogue infrastructure, or insider privileges -- can intercept these measurements and covertly track the movement of people and robots throughout the facility.

Such leakage exposes highly sensitive information. For passengers, it reveals destinations, dwell times, and behavioral patterns; for service robots and security drones, it exposes patrol routes and operational routines, undermining their effectiveness. These insights can be exploited for a spectrum of attacks, from targeted manipulation and behavioral profiling to operational disruption and espionage. Importantly, even if each stakeholder (airlines, retailers, airport operations, security) maintains separate IT systems, all localization traffic ultimately traverses the same underlying network provider (e.g., AT\&T, Verizon). This shared infrastructure forms a single point of compromise, making secure, privacy-preserving indoor wireless localization essential.

In \name, our federated architecture prohibits such attacks through a change in paradigm -- where the user is back in control of estimating their own location. When a user requests their location, the user's device pings each AP to collect CSI data. Crucially, the CSI data is processed into AoA-ToF images and encoded \textbf{locally at each AP}, producing a compact embedding. Each AP's embedding is then sent directly to the user device with no intermediary, where the user's local decoder aggregates the embeddings to produce a final location estimate. This design ensures that raw CSI data never leaves the AP, fundamentally limiting privacy exposure while maintaining localization accuracy. Furthermore, even if an attacker compromises individual APs, they can only acquire high-dimensional, complex-valued raw CSI data or encoded embeddings -- neither of which reveal location information without the corresponding decoder held privately by the user.


\subsection{Motivation for AoA-Based Localization}

Prior works \cite{dloc} have demonstrated impressive results using XY-coordinate-based images as input to their models. These XY-images, also generated from CSI data, have distinct advantages: they represent CSI measurements over a global 2D coordinate plane and inherently encode AP locations within the image structure, providing a straightforward representation for neural networks to estimate user location with minimal coordinate transformations. However, this approach has a critical limitation: image dimensions scale with environment size, preventing true generalization across different-sized environments. A naive solution would be to apply padding to the images to achieve uniform dimensions. However, this fails to scale for real-world deployment where floor plans can span thousands of square feet, resulting in prohibitively large images that cannot be processed in real-time. In contrast ~\cite{zhang2024rloc} relies on AoA measurements across successive timestamps to estimate accurate AoA but also provides more attack points for the user location privacy.

\begin{figure*}[!ht]
    \centering
    \includegraphics[width=\linewidth]{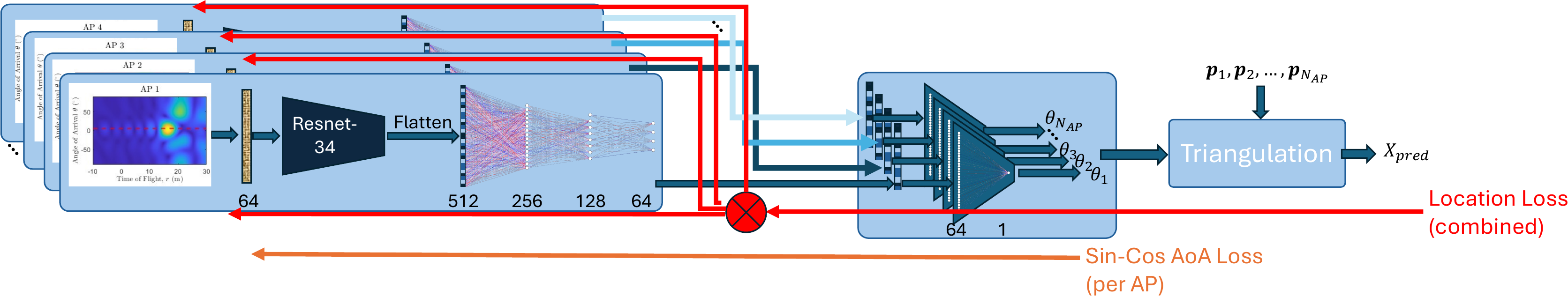}
    \caption{Model Architecture:\normalfont{ Shows the encoder-decoder architecture of \name. We use Resnet-34 modules for the encoder~\cite{resnet-pt}} and the decoder receives the $N_{AP}$ embeddings and predicts $\{\theta_1,\theta_2,\cdots,\theta_{N_{AP}}\}$ that are then used along with $\{\mathbf{p}_1,\mathbf{p}_2,\cdots,\mathbf{p}_{N_{AP}}\}$ to perform triangulation. The triangulation loss is combined till the shared decoder but splits to propagate appropriate gradients through each AP's Encoder. The $\sin(\theta)$ and $\cos(\theta)$ losses are independent for each AP.}
    \label{fig:model_architecture}
\end{figure*}

Thus, to achieve cross-environment generalization, we adopt the idea of estimating accurate AoAs but only from a single snapshot of CSI, instead of getting a distribution over successive timestamps. We use AoA-ToF image representation of CSI data, which generates heatmap-like visualizations of users relative to each AP. The representation offers two key advantages:

\begin{enumerate}[leftmargin=*]
    \item \textbf{Fixed Dimensions}: Image size remains constant regardless of environment dimensions, ensuring computational uniformity across deployments
    \item \textbf{Rich Multipath Representation}: The images capture both direct-path and multipath propagation characteristics with high fidelity
\end{enumerate}

This representation also simplifies training across diverse datasets. Provided the model is trained over the full angular range, it can generalize to multiple environments with varying multipath characteristics because the network only needs to learn two fundamental tasks:

\begin{enumerate}[leftmargin=*]
    \item \textbf{LOS Scenarios}: Accurately predict the angle from the dominant LOS peak, accounting for small offsets
    \item \textbf{NLOS Scenarios}: Distinguish between LOS and multipath peaks to identify the true direct path
\end{enumerate}
Critically, the network does not need to learn inter-AP relationships, AP locations, or environment-specific metadata, making this architecture inherently generalizable. AP locations do need to be known for the final AoA-to-XY transformation. However, that is a minor computation and can be performed reliably on the user's devices with very minimal compute overhead.

\section{Design}\label{sec:methodology}

\subsection{Model Architecture}

Our system (Fig.~\ref{fig:model_architecture}) employs a distributed architecture that prioritizes privacy and computational efficiency through decentralized processing at Access Points (APs). Rather than transmitting raw location-sensitive data, each AP processes AoA-ToF images locally using a ResNet34 encoder (\S\ref{sec:encoder}) and transmits only a compact 64-dimensional embedding vectors to the target device. The target device aggregates embeddings from multiple APs and processes them through a multilayer perceptron (MLP) decoder (\S\ref{sec:decoder}) that predicts AoA values. The predicted trigonometric components are transformed to global coordinates through a least-squares triangulation algorithm (\S\ref{sec:triang}).

Lastly, our system is trained using a federated architecture where every 10 batches, the encoders are all averaged. During deployment, each AP does not retrain its own encoder. Rather, the averaged encoder will be given to the AP to generate its location embedding vectors, as described in section \S\ref{sec:federated}.

\subsubsection{Encoder Design}\label{sec:encoder}
The encoder faces a fundamental challenge: extracting robust directional features from AoA-ToF images while suppressing multipath interference. In indoor environments, signal reflections create spurious peaks that can mislead conventional feature extraction methods. 

We selected ResNet34, originally developed for image recognition, as our encoder backbone. The deep residual pathways used for retaining spatial information across convolutional layers enables the network to differentiate line-of-sight (LOS) from multipath peaks. Shallow CNNs may filter out the LOS peak in midst of a stronger multi-path peak which can lead to poor generalization in multipath-rich environments. Our choice of ResNet34 over deeper variants (ResNet50, ResNet101) reflects practical deployment constraints. Since encoders must operate on resource-constrained Access Points, we prioritized computational efficiency while maintaining sufficient representational capacity. ResNet34 provides an optimal balance between model complexity and edge device compatibility.

We adapted the standard ResNet34 architecture with two key modifications: (1) the initial 3-channel convolutional layer is replaced with a single-channel variant to process grayscale AoA-ToF images, and (2) the classification head is substituted with a multi-layer perceptron that compresses high-dimensional convolutional features into 64-dimensional embeddings. 

The multi-layer perceptron comprises three linear layers each followed by ReLU activation and dropout regularization layers. The final layer outputs a 64-dimensional embedding which was chosen to balance information preservation with communication efficiency. At 256 bytes per embedding (64 float32 values × 4 bytes each), the payload remains extremely lightweight, critical for real-time localization applications where multiple APs must transmit simultaneously to the target device.


\subsubsection{Decoder Design}\label{sec:decoder}

The decoder must accurately predict angle-of-arrival values from 64-dimensional embeddings across diverse environments and AP configurations. While a simple MLP with mean squared error (MSE) loss suffices for single-environment deployment, generalization presents significant challenges: APs in different environments exhibit varying degrees of multipath propagation, interference, and path loss, requiring the decoder to be robust to these environmental variations.

We explored several architectural approaches before arriving at our final design. First, we attempted to leverage temporal smoothness between consecutive CSI samples using recurrent architectures including Gated Recurrent Units (GRUs) and Long Short-Term Memory (LSTM) networks. While these models performed well in single environments, they failed to generalize across multiple environments since angle paths may not always be smooth across scenarios with varying levels of multi-path and interference. Next, we developed a dual-branch MLP decoder that independently predicted $\sin(\theta)$ and $\cos(\theta)$ components. This architecture showed promise in both single and multi-environment settings but proved unnecessarily complex for the task, under-performing our final design in cross-environment evaluation.

Our final decoder comprises a series of linear layers with ReLU activation and dropout regularization, terminating in a hyperbolic tangent ($\tanh$) activation function. The $\tanh$ function constrains outputs to $[-1, 1]$, which are then scaled by $\pi/2$ to produce angle-of-arrival predictions in radians. Despite its simplicity, this architecture, when combined with our geometric loss function (\S\ref{sec:loss}), achieves strong generalization across environments while maintaining excellent single-domain performance. The key insight is that direct angle regression with appropriate geometric constraints proves more effective than complex temporal or trigonometric decompositions for cross-environment localization.

\subsubsection{Triangulation}\label{sec:triang}

Target localization employs least-squares triangulation to resolve bearing measurements from multiple APs into a single position estimate. Each AP contributes a ray that can be formulated as $\mathbf{r}(t) = \mathbf{p} + t \mathbf{d}$, where $\mathbf{p} = (x,y)$ is the location of the AP, and $\mathbf{d} = (\cos(\theta), \sin(\theta))$ is the unit vector indicating the ray's direction, determined by the bearing $\theta$ (AoA) to the target. The $\cos(\theta)$ and $\sin(\theta)$ values used in $\mathbf{d}$ are the output of the MLP decoder, transformed into the global coordinate frame. Due to measurement inaccuracies, these lines usually do not intersect at a single point, necessitating a least-squares approach to find the geometrically optimal central point that best fits all the noisy bearing observations \cite{ray}.

The target location $\mathbf{(x,y)}$ is the least-squares solution to the following linear system.

$$
\begin{bmatrix}
\sum_{i}(1 - a_i^2) & -\sum_{i} a_i b_i \\[1em]
-\sum_{i} a_i b_i & \sum_{i}(1 - b_i^2)
\end{bmatrix}
\begin{bmatrix}
\mathbf{x} \\[1em]
\mathbf{y}
\end{bmatrix}
=
\begin{bmatrix}
\sum_{i}[(1 - a_i^2)x_i - a_i b_i y_i] \\[1em]
\sum_{i}[(1 - b_i^2) y_i -a_i b_i x_i]
\end{bmatrix}
$$

where $a_i = \cos(\theta_i), \space b_i= \sin(\theta_i)$. $\theta_i$ is the AoA measurement from the $i$-th AP, and $(x_i,y_i)$ is the location of the $i$-th AP.

\subsubsection{Geometric Loss Function}\label{sec:loss}

Traditional loss functions such as MSE or L1 loss are inadequate for direct angle prediction in our system. They fail to handle angular periodicity and create disproportionate penalties near $\pm \pi/2$, where antenna geometry creates inherent directional ambiguity. To address this, we introduce a geometric loss function that decomposes angles into trigonometric components:

\begin{align*}
\mathcal{L}_{\text{total}} &= \lambda_1 \cdot \text{Huber}(\cos(\theta_{\text{gt}}), \cos(\theta_{\text{pred}}), \delta_1) \\
&+ \lambda_2 \cdot \text{Huber}(\sin(\theta_{\text{gt}}), \sin(\theta_{\text{pred}}), \delta_2) \\
&+ \lambda_3 \cdot \text{Huber}(\mathbf{X}_{\text{gt}}, \mathbf{X}_{\text{pred}}, \delta_3)
\end{align*}

where $\theta$ represents the angle-of-arrival, $\mathbf{X}$ is the 2D location obtained via triangulation (\S\ref{sec:triang}), $\delta_i$ controls the Huber threshold between quadratic and linear regions. We set $\delta_1=0.5$, $\delta_2=1.0$, $\delta_3=2.0$ based on empirical tuning. The location term enforces geometric consistency across multiple APs, ensuring that predicted angles yield physically plausible positions.

Both sine and cosine components are essential because they provide complementary sensitivity across the angular range. The gradient of sine loss is proportional to $\cos(\theta)$, which vanishes near $\pm 90°$, while cosine loss has gradients proportional to $\sin(\theta)$, which vanishes near $0°$. Using both ensures no angular region suffers from vanishing gradients. Consider two illustrative cases that demonstrate this complementarity:

\begin{itemize}[leftmargin=*]
    \item \textbf{Ambiguous vertical angles:} Suppose $\theta_{\text{gt}} = 88°$ but the model predicts $\theta_{\text{pred}} = -85°$ (a 173° error that appears similar in the AoA-ToF image due to top/bottom edge peaks). The cosine values are nearly identical: $\cos(88°) = 0.035 \approx \cos(-85°) = 0.087$, yielding minimal loss ($\approx 0.05$). However, the sine component captures this failure: $|\sin(88°) - \sin(-85°)| = |0.999 - (-0.996)| \approx 2.0$, providing strong corrective feedback.
    
    \item \textbf{Small errors near broadside:} For $\theta_{\text{gt}} = 5°$ and $\theta_{\text{pred}} = 25°$ (a 20° error), sine provides good sensitivity since it's approximately linear near zero: $|\sin(5°) - \sin(25°)| \approx |0.087 - 0.423| = 0.336$. The cosine component also contributes: $|\cos(5°) - \cos(25°)| \approx |0.996 - 0.906| = 0.090$. Both losses work together to correct the prediction.
\end{itemize}

Using sine alone would be insufficient despite eliminating $\pm 90°$ ambiguity: sine's gradient $\cos(\theta)$ approaches zero near vertical angles, causing slow convergence in these regions. This creates biased training where horizontal angles learn faster than vertical ones. The cosine term compensates precisely where sine weakens, ensuring uniform learning across all angles.

We employ Huber loss rather than MSE for its robustness to outliers. In multipath-rich environments where spurious peaks can cause large prediction errors, Huber loss transitions from quadratic (for small errors) to linear (for large errors), preventing outliers from dominating the gradient. The $\delta$ parameters control this transition point, with values selected through systematic hyperparameter search detailed in \S\ref{sec:ablation}.
\subsection{Federated Learning} \label{sec:federated}
\subsubsection{Privacy Risk and Motivation} 
Our system faces privacy risks from two primary threat vectors: (1) external attackers who may compromise individual Access Points (APs) or intercept network communications, and (2) honest-but-curious infrastructure operators who have legitimate access to system components but may attempt to infer user locations. In a centralized approach, compromising a single server would expose all user location data and movement patterns. To mitigate these risks, we propose a federated learning framework that decentralizes both training data and inference computation.
\begin{figure}[t]
    \centering
    \includegraphics[width=0.9\linewidth]{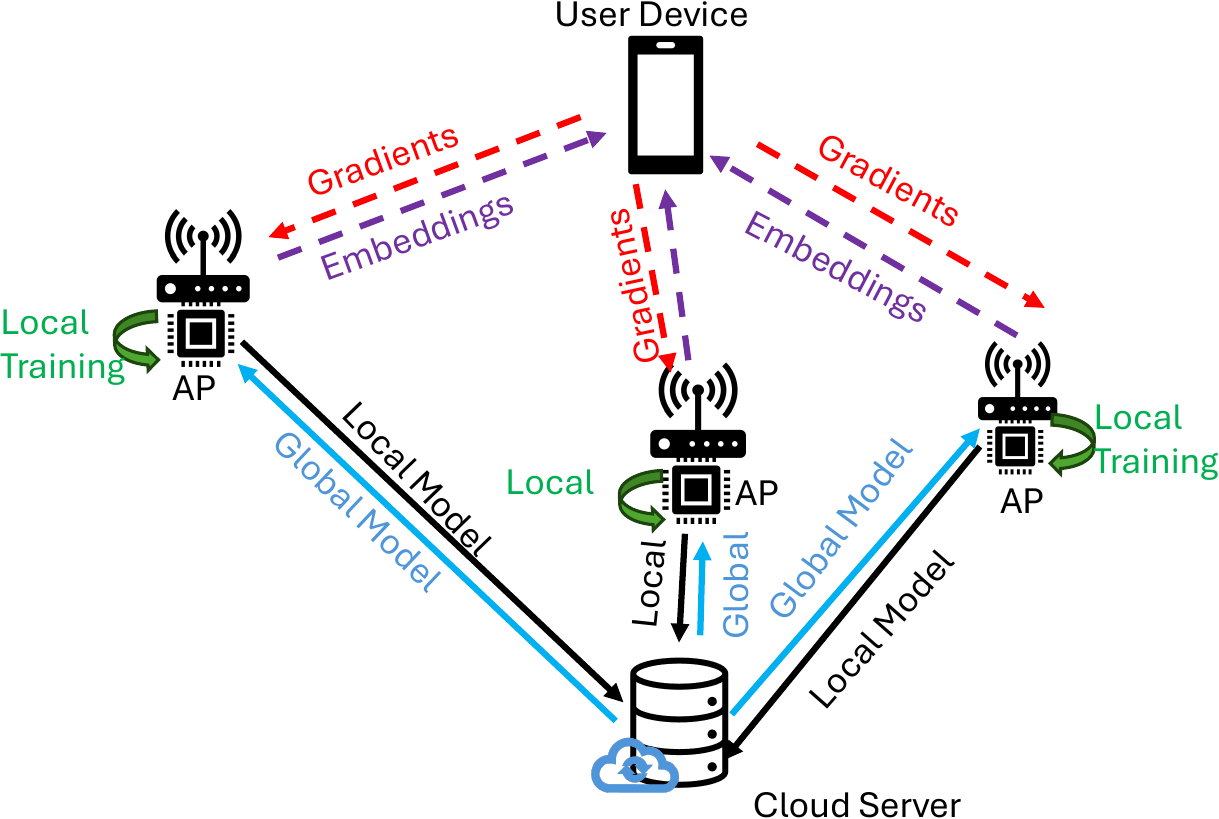}
    \caption{Federated Learning}
    \label{fig:fed_learning}
\end{figure}

\subsubsection{Federated Training Architecture}
In our federated framework (Fig.~\ref{fig:fed_learning}), local training nodes independently train models on their respective localized datasets. Rather than transmitting raw data to a central cloud server, each local server securely sends only its trained model to the cloud. The central cloud server then aggregates these local models into a global model using the \textbf{FedAvg} algorithm \cite{fedavg}. The central server aggregates these parameters into a global model and redistributes the updated weights back to local servers.

This approach addresses the geographic data distribution challenge inherent in localization: while each local server trains primarily on data from its region, the federated aggregation enables the global model to learn representations that generalize across different environments and AP configurations. The iterative training process allows the model to benefit from diverse deployment scenarios while keeping location-sensitive data localized.

\subsubsection{Privacy-Preserving Deployment}
During deployment, each AP runs a local encoder module while users maintain the decoder on their devices. When a location request is initiated, participating APs (typically 3-4 for triangulation) simultaneously collect CSI measurements, process them through their local encoders, and transmit 64-dimensional embedding vectors to the requesting user. The user's device aggregates these embeddings, and performs triangulation via the decoder
for the final location prediction.

\subsubsection{Privacy Analysis}
This architecture provides several privacy benefits compared to a typical centralized approach:
\begin{enumerate}[leftmargin=*]
    \item \textbf{Data Localization:} Raw CSI data never leaves the AP, limiting the exposure to detailed channel information
    \item \textbf{Reduced Attack Surface:} Compromising a single AP only reveals a subset of the data and contains no information from nearby APs
    \item \textbf{Computation Distribution:} Final location estimations are done on the user's device, preventing man-in-the-middle attacks from intercepting the location of the user
    \item \textbf{Embedding Obfuscation:} While the 64-dimensional vector embeddings can be compromised during transmission from the AP to the user, they require the corresponding decoder to extract useful information. 
\end{enumerate}
Although attackers still have opportunities to compromise individual APs or intercept the vector embedding from the AP to the user, the federated architecture significantly raises the attack complexity compared to a centralized approach for the ability to extract useful information during any stage of the localization process.

\section{Implementation}\label{sec:implementation}
This section details our implementation choices, training procedures, and dataset pre-processing steps necessary for reproducible results. Our open-sourced code base can be found here with links to the datasets: https://github.com/WIRES-UB/PriWiLoc-Open. We also include all data pre-processing scripts in the code base. 

\begin{table*}[t]
  \caption{Overview of Evaluation Datasets}
  \centering
  
  \begin{tabular}{|c|c|c|c|c|p{7cm}|}
    \hline
    \textbf{Abbr.} & \textbf{Scenario} & \textbf{Location} & \textbf{Space Size} & \textbf{No. of APs} & \textbf{Description}  \\ \hline 
    E1 & July 28th & Jacobs Hall & 18m $\times$ 8m & 4 & Complex high-multipath and NLOS environment (1500 sq. ft.).  \\ \hline
    E2.1 & Aug. 16th-1 & Jacobs Hall & 18m $\times$ 8m & 4 & Complex high-multipath and NLOS environment (1500 sq. ft.), with extra furniture placed randomly.  \\ \hline
    E2.2 & Aug. 16th-3 & Jacobs Hall & 18m $\times$ 8m & 4 & Complex high-multipath and NLOS environment (1500 sq. ft.), with extra furniture placed randomly.  \\ \hline
    E2.3 & Aug. 16th-4 & Jacobs Hall & 18m $\times$ 8m & 4 & Complex high-multipath and NLOS environment (1500 sq. ft.),  an added reflector. \\ \hline
    E3 & Conference & RLoc-Conference & 8m $\times$ 8m & 4 & LoS-based environment in conference room. \\ \hline
    E4 & Lab & RLoc-Lab & 9m $\times$ 10m & 4 & LoS-based environment in laboratory. \\ \hline
    E5 & Office & RLoc-Office & 9m $\times$ 11m & 4 & Complex high-multipath environment in office. \\ \hline
    E6 & Lounge & RLoc-Lounge & 11m $\times$ 14m & 4 & Complex high-multipath and NLOS environment in lounge. \\ \hline
  \end{tabular}
  \label{tab:dataset-overview}
\end{table*}

\subsection{System Implementation}
We implemented our system completely using PyTorch \cite{paszke2019pytorch} and PyTorch Lightning \cite{Falcon_PyTorch_Lightning_2019}. For the encoder, we used torchvision's implementation of ResNet34. We modified the ResNet34 architecture in two key areas: (1) input layer adaptation for multi-AP data, and (2) output layer redesign for dimensionality reduction (Fig.~\ref{fig:model_architecture}). At the input layer, we adjust the \textit{in\_channels} parameter of the \textit{Conv2d} block to take in the number of APs. For the final layer, we remove the final classification layer in exchange for \textit{nn.Identity()}. The result from the output layer is then passed through a Multilayer Perceptron (Fig.~\ref{fig:model_architecture}) to reduce the output dimension. For the decoder, it is implemented in the following order: a \textit{Linear} layer (input features → 256), \textit{ReLU} activation, \textit{Dropout} $(p=0.2)$, a second \textit{Linear} layer (256 → $n_{ap}$), and \textit{Tanh} activation to produce the AoA predictions.

For federated learning, we created a separate class that inherits from the base model class. The federated class takes care of the parameter averaging. It extracts all the layers within the encoder module, averages all except the batch norm layers, and returns the parameters for each encoder to use. Through experimental means, we found that averaging every $10$ batches gives the best performance in terms of location prediction accuracy while maintaining privacy. This split-style architecture allows us to utilize all the functionality given with PyTorch Lightning while adding our custom modules. 

\subsection{Hyperparameter Tuning and Training}
Through extensive evaluation, we found these parameters to be optimal for most training scenarios. We use the Adam Optimizer \cite{adam_optimizer} with a default learning rate of $5 * 10^{-5}$. We found the \textit{OneCycleLR} scheduler to be beneficial 
in converging quicker. The learning rate scheduler updates every step and has a maximum learning rate of $5 * 10^{-5}$. We found batch size of 32 and 100 epochs is sufficient for training.

\subsection{Datasets}
To ensure our model can be validated against any state-of-the-art systems, we used the datasets found in both the DLoc \cite{dloc} and RLoc \cite{zhang2024rloc} papers. The datasets comprises of $\sim$100,000 WiFi CSI measurements gathered across 6 different environments totaling over 5000 sq. ft., and across two different bandwidths (80Mhz, 40Mhz).  The environments include both simple (600 sq. ft.) and complex (1,500 sq. ft.) spaces with rich multipath propagation, non-line-of-sight conditions, and various furniture configurations. In DLoc datasets, ground truth localization was provided through simultaneous SLAM mapping, with CSI measurements sampled every 50ms from 3-4 access points per environment. In RLoc, they use ultra-wideband localization systems with an accuracy of ~10cm with millisecond-level synchronization.

We pre-process this CSI data in MATLAB to create individual \textbf{HDF5} files for each measurement. We transform the CSI Data to the AoA-ToF format as our input data for the model. The transformation from CSI to AoA-ToF images is described in these past works \cite{dloc, kotaru2015spotfi, deepwifi_csi_net, Halperin2011CSItool}. Each file comprises of an AoA-ToF image for each AP, ground truth AoA for each AP, ground truth location in $(x, y)$ coordinates, velocity measurement, measurement timestamp, and received signal strength indicator for each AP. This format allows us to train on various sections of different environments to understand where our model may come short. 

\section{Evaluation}\label{sec:Evaluation}
In this section, we do a thorough end-to-end evaluation, testing our system against multiple datasets and state-of-the-art Wi-Fi localization systems. We also go through important ablation studies, showcasing the importance of each of the components in our system. 
\begin{figure}[t]
    \centering
    \includegraphics[width=0.9\linewidth]{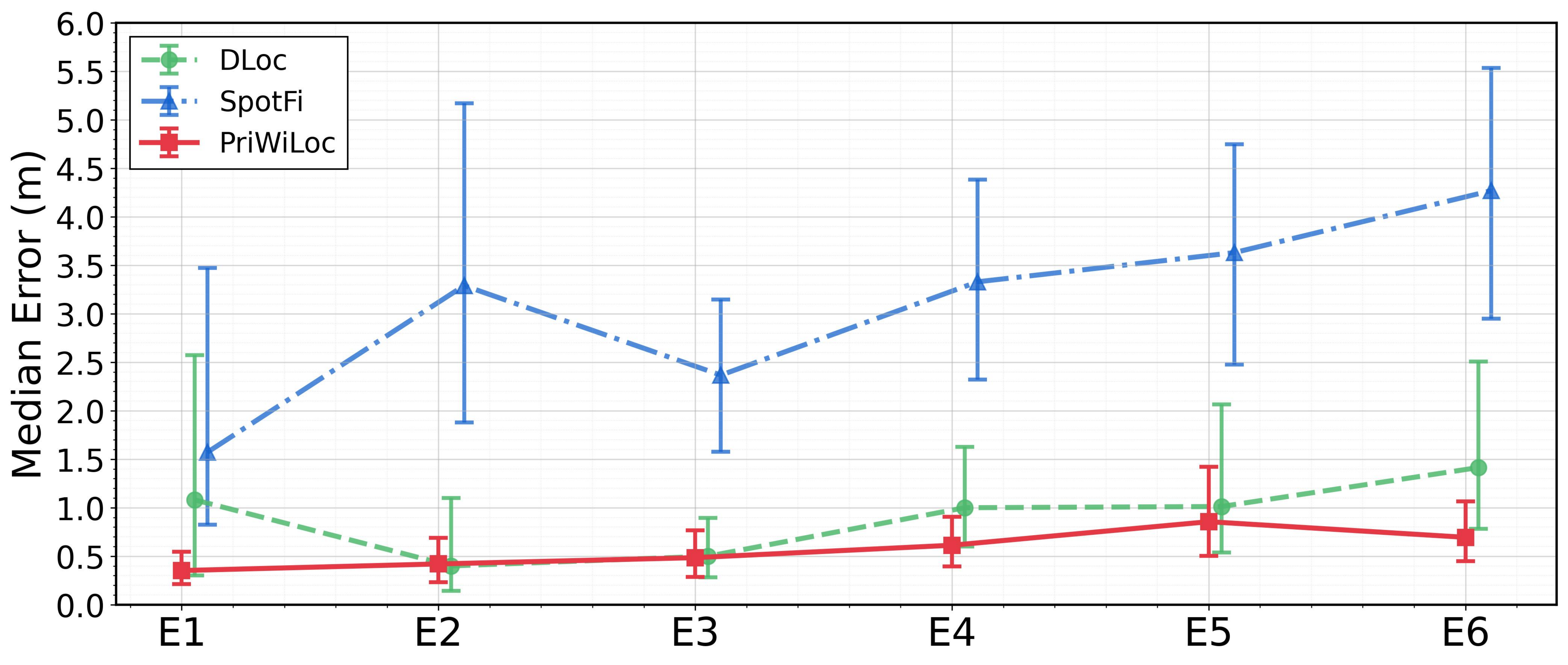}
    \caption{\name Results: \normalfont \name outperforms state-of-the-art models (DLoc, Spotfi) in localization in multiple environments of varying sizes}
    \label{fig:core_test}
\end{figure}

\begin{table}[t]
\centering
\caption{90th$\%$ile Error (m) for In-domain experiments}
\begin{tabular}{lccc}
\hline
Env & DLoc & PriWiLoc & SpotFi \\
\hline
E1 & 5.572 & 0.776 & 6.115 \\
E2 & 2.625 & 1.072 & 7.355 \\
E3 & 2.502 & 1.188 & 3.914 \\
E4 & 2.693 & 1.277 & 5.413 \\
E5 & 4.308 & 2.184 & 5.868 \\
E6 & 4.319 & 1.609 & 6.840 \\
\hline
\end{tabular}
\label{tab:90th_percentile_error}
\vspace{-2em}
\end{table}

\begin{figure*}[t]
        \centering
        \includegraphics[width=0.98\linewidth]{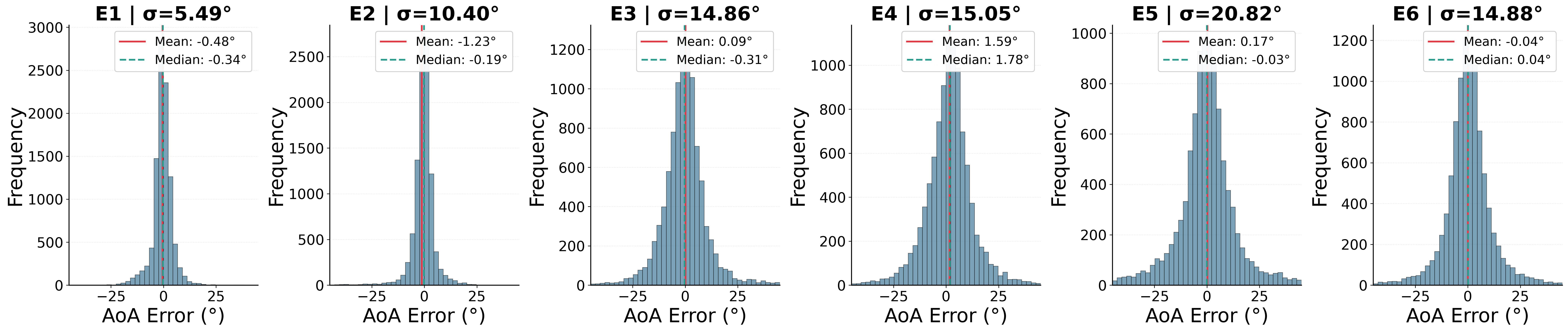}
    \caption{AoA errors across various environments: Histograms of AoA error (Ground truth - Predicted) distributions across all the 6 different environments \name has been tested against}
    \label{fig:aoa_vs_pred}
\end{figure*}
\subsection{End-to-End Evaluation}
For our benchmarks, we compare our system with Spotfi \cite{kotaru2015spotfi}, and DLoc \cite{dloc}. Spotfi utilizes a super-resolution algorithm to simultaneously predict the AoA and ToF with high accuracy. Note that there is no open-source implementations of RLoc\cite{zhang2024rloc} so we are unable to compare directly, but we do run \name on their datasets. DLoc uses a CNN image-to-image translation network to model the environment and predict a location prediction heatmap. For single-environment scenarios, we train on 80\% and test on 20\% of the dataset. For cross-environment scenarios, we train on multiple environments and test on a new environment.

\subsubsection{In-Domain Localization Accuracy --}
Figure~\ref{fig:core_test} presents median localization error with 25th and 75th percentile markers across multiple environments, comparing \name against DLoc and Spotfi. \name consistently outperforms Spotfi across all environments, achieving an 81\% average reduction in median error. Compared to DLoc, \name achieves an average improvement of 28.5\% in median error, with performance ranging from a 61.9\% improvement (65 cm) in scenario E1 to a slight 12.5\% degradation (5 cm) in scenario E2. Notably, \name demonstrates tighter error distributions with an average median error of 58 cm and average 90th percentile of 135 cm, compared to 366 cm and 591 cm for DLoc and Spotfi, respectively.

\begin{figure*}[t]
    \centering
    \begin{subfigure}[b]{0.8\linewidth}
        \centering
        \includegraphics[width=1.0\linewidth]{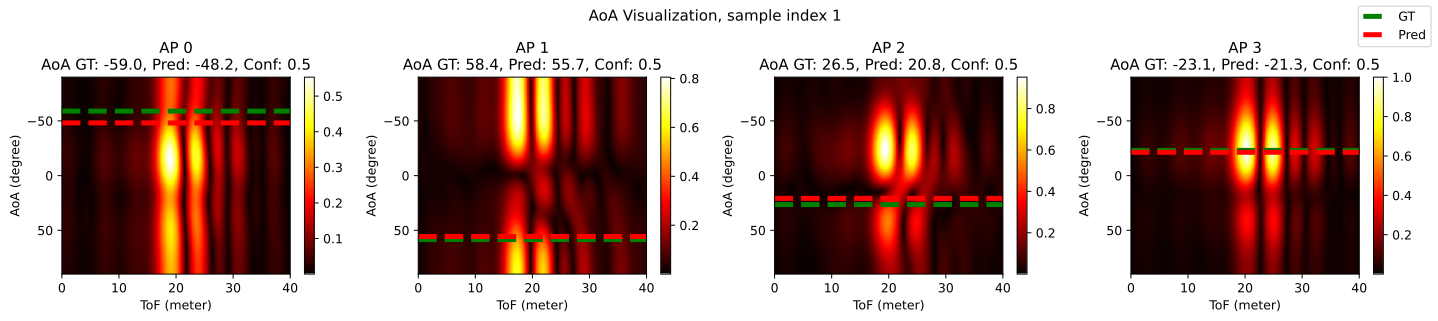}
        \caption{\normalfont 80MHz July 28th Dataset}
    \end{subfigure}
    \vspace{0.1cm} 
    \begin{subfigure}[b]{0.8\linewidth}
        \centering
        \includegraphics[width=1.0\linewidth]{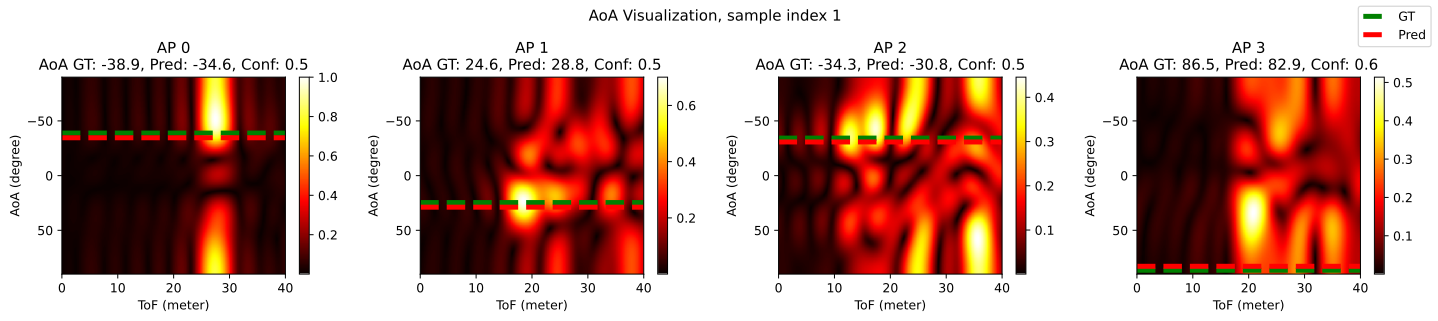}
        \caption{\normalfont 40MHz Conference Dataset}
    \end{subfigure}
    \vspace{-1em}
    \caption{AoA-ToF Images for Each AP with Ground Truth and Predicted AoA}
    \label{fig:aoa-tof-image_pred}
\end{figure*}
\subsubsection{Robust and Accurate AoA Prediction --}
In figure~\ref{fig:aoa_vs_pred}, we present the AoA error distributions across all six environments, demonstrating \name's robust angle estimation capabilities. Across all scenarios, \name achieves a maximum mean AoA error of only 1.59$^\circ$, with five scenarios achieving sub-degree median error. Scenarios E3-E6 exhibit higher standard deviations due to the challenging 40 MHz data, where strong multipath effects make consistent angle prediction more difficult. In contrast, with 80 MHz data (E1 and E2), standard deviations drop to 5.49$^\circ$ and 10.40$^\circ$ respectively, demonstrating \name's ability to handle interference, obstructions, and NLOS conditions when sufficient bandwidth is available.

To demonstrate \name's robustness to multipath effects, Figure~\ref{fig:aoa-tof-image_pred} visualizes AoA-ToF images for each AP alongside ground truth and predicted AoA values across two environments: 80 MHz (E1, July 28th) and 40 MHz (E3, Conference). The figure illustrates the inherent difficulty in processing both bandwidths—40 MHz data exhibits wider peaks and stronger multipath components, making direct path identification more challenging.
In scenario E1, \name achieves AoA errors ranging from 1.8$^\circ$ to 8.8$^\circ$ for the displayed sample. Notably, in AP 2, a strong multipath peak could easily be mistaken for the direct path, yet \name correctly identifies the weaker ground truth peak. For scenario E3, AoA errors range from 3.5$^\circ$ to 4.3$^\circ$. AP 3 exemplifies the challenge of 40 MHz data, where multiple strong peaks complicate direct path identification.
This is where geometric loss proves critical -- by jointly optimizing angles with localization error, the model learns which angle combinations are geometrically plausible. Since even a single erroneous angle can cause large triangulation errors and high localization loss, this constraint effectively teaches the model to distinguish direct path from multipath peaks across all APs simultaneously.

\begin{table*}
    
    \caption{Generalization results across changing environments: Median and 90th percentile errors when trained and tested on across different setups of DLoc datasets.}
    \centering
        \begin{tabular}{|c|c|c|c|c|c|c|c|}
        \hline
        \multirow{2}{*}{Trained} & \multirow{2}{*}{Tested} & \multicolumn{3}{c|}{Median Error (cm)} & \multicolumn{3}{c|}{$90^{th}\%$ile Error (cm)} \\
        \cline{3-8}
        & & \name & DLoc & SpotFi & \name & DLoc & SpotFi \\
        \hline
        E1,E2.2,E2.3 & E2.1 & 86 & 120 & 198 & 239 & 310 & 420 \\
        \hline
        E1,E2.1,E2.3 & E2.2 & 63 & 130 & 154 & 213 & 380 & 380 \\
        \hline
        E1,E2.1,E2.2 & E2.3 & 134 & 120 & 161 & 256 & 310 & 455 \\
        \hline
        \end{tabular}
        \label{tab:cross-scenario-table}
\end{table*}

\subsection{\name's Generalization Performance}
To fully evaluate \name's capabilities and improvements over state-of-the-art models such as DLoc \cite{dloc} and SpotFi \cite{kotaru2015spotfi}, we examine multiple cross-environment generalizations scenarios.

\begin{figure}[t]
\centering
    \includegraphics[width=1.0\linewidth]{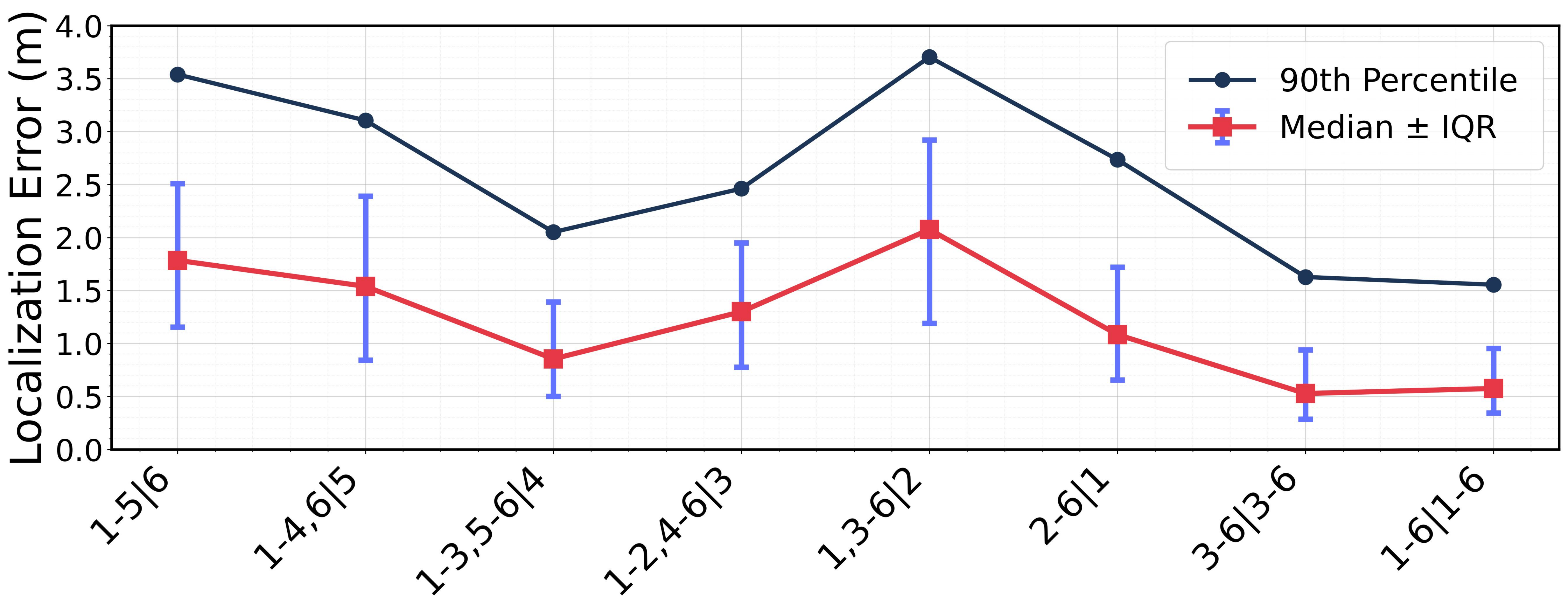}
    \caption{Median and 90th$\%$ile Localization Error  for cross-scenario experiments ${X | Y}$ where $X$ represents the training scenarios and $Y$ the test scenario(s)}
    \label{fig:generalization}
\end{figure}
\begin{figure*}[ht]
    \centering
    \begin{minipage}{0.33\linewidth}
    \includegraphics[width=1\linewidth]{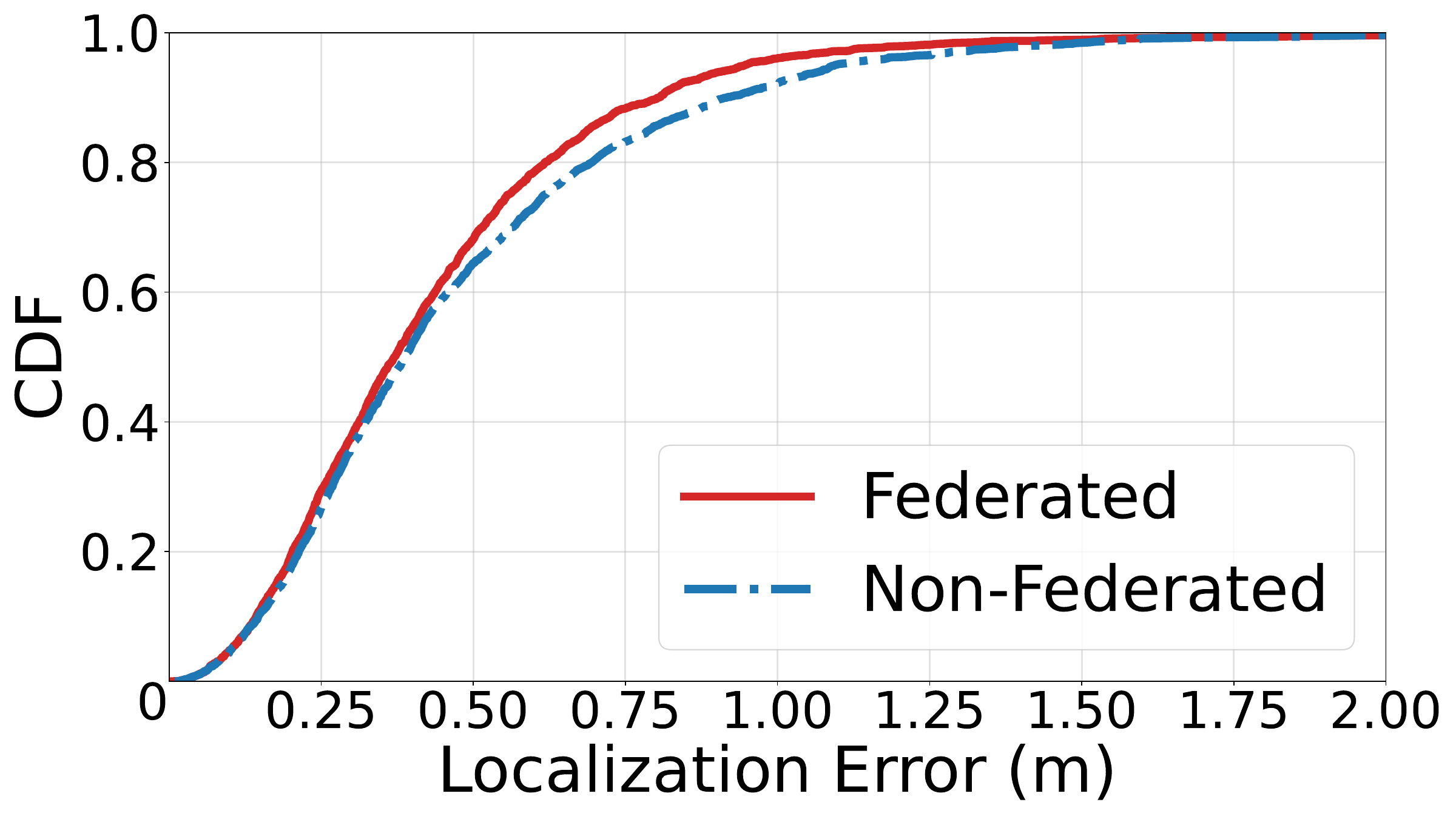}
    \subcaption{}
    \end{minipage}
    \begin{minipage}{0.33\linewidth}
    \includegraphics[width=1\linewidth]{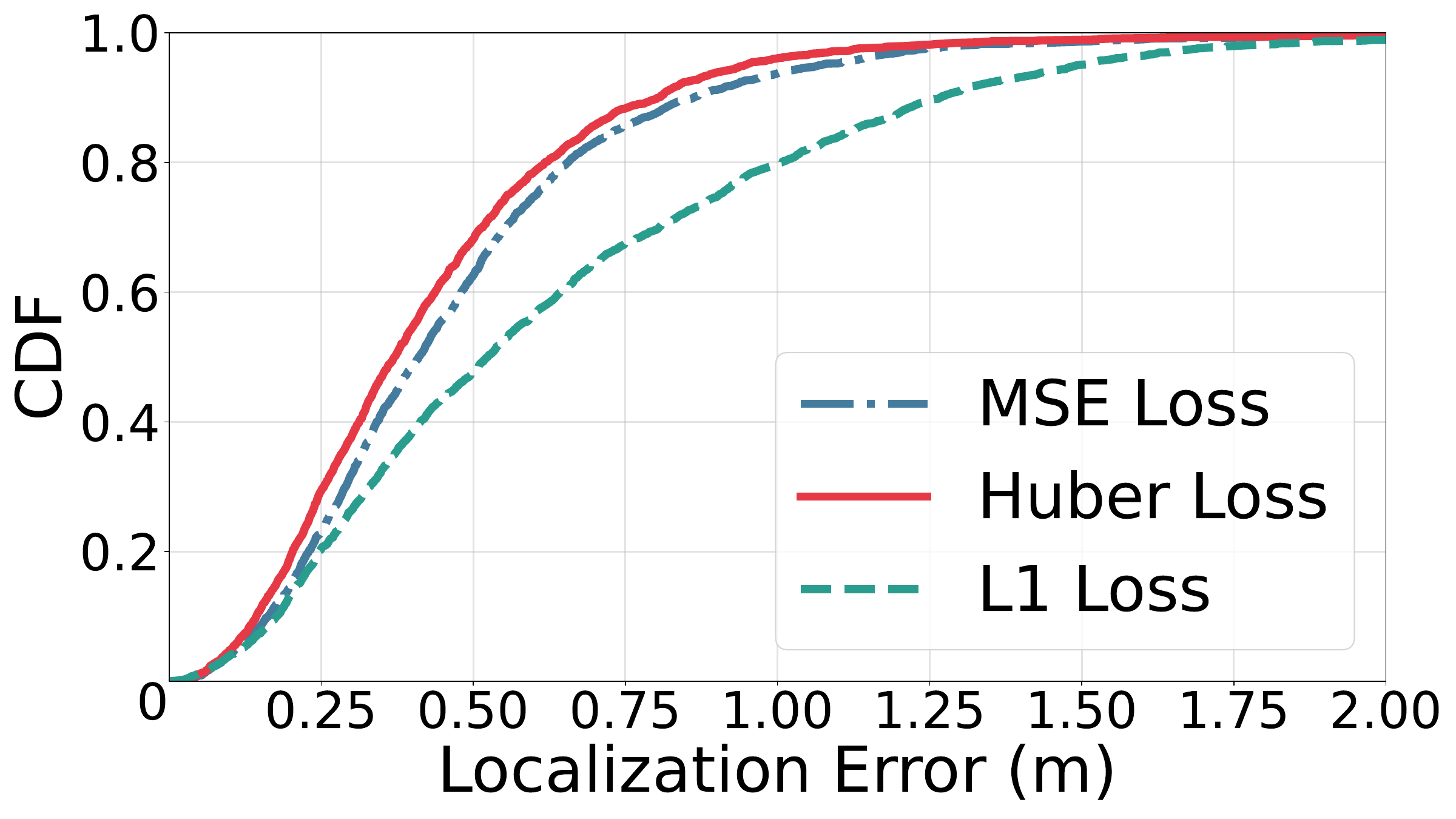}
    \subcaption{}
    \end{minipage}
    \begin{minipage}{0.33\linewidth}
    \includegraphics[width=1\linewidth]{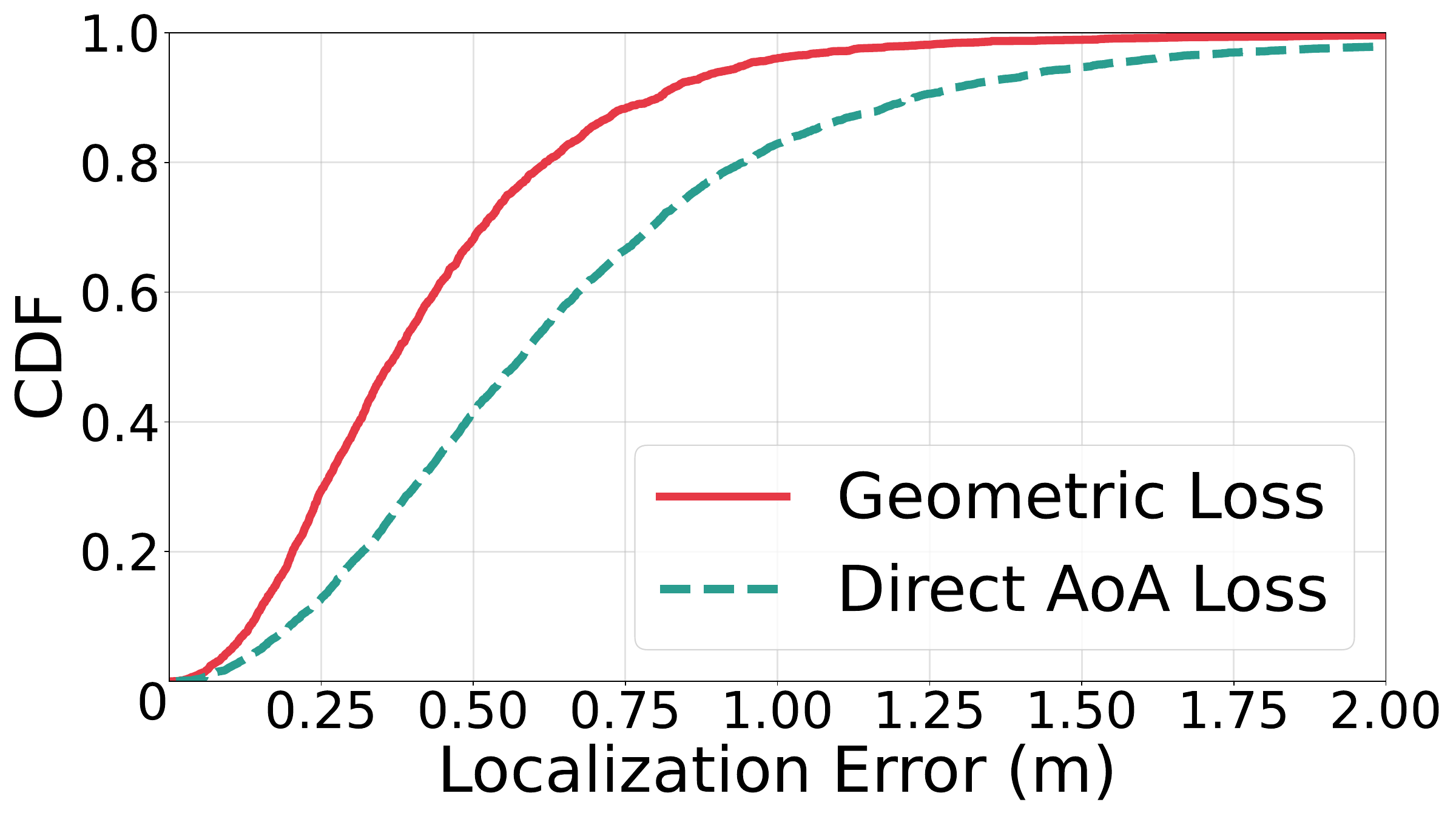}
    \subcaption{}
    \end{minipage}
    \vspace{-1em}
    \caption{Ablation Study: (a) \normalfont{Federated learning provides minor improvement over non-federated learning while providing a privacy component} (b): \normalfont{Huber Loss provides large improvement over L1-Loss and closely matches the performance of MSELoss with some improvement} (c) \normalfont{Geometric Loss noticeably outperforms Direct-AoA Loss across all metrics}}
    \label{fig:ablation}
\end{figure*}

\subsubsection{Generalization Across Changing Spaces --}
We first examine Table~\ref{tab:cross-scenario-table}, where the model is trained on three scenarios from the same general area with slight modifications and tested on a fourth unseen scenario. The scenarios include E1, E2.1, E2.2, and E2.3. This experiment demonstrates \name's ability to accurately localize users despite environmental changes that introduce interference, multipath effects, and potential non-line-of-sight (NLOS) conditions.

In the first test scenario, \name achieves a median localization error 34 cm lower than DLoc and 112 cm lower than SpotFi. In scenario 2, we are able to achieve a 67 cm improvement over DLoc and 91 cm improvement over Spotfi, demonstrating robustness to random furniture rearrangements that create new reflections and multipath effects.  Scenario 3 introduces an aluminum reflector that produces significant multipath interference. While DLoc achieves a slightly lower median error (14 cm improvement), \name demonstrates superior consistency with a 54 cm improvement in 90th percentile error. Notably, \name's 90th percentile error remains consistently below 300 cm across all scenarios, whereas DLoc and SpotFi exhibit significantly higher variability, reinforcing that \name is able to generalize to new scenarios with more accuracy than the state-of-the-art. 

\subsubsection{Cross-Domain Generalization --}

To the best of our knowledge, \name is the first system capable of localizing across both different environments \textbf{and} different bandwidths using a single model. DLoc cannot generalize across environments with the same model due to their reliance on XY-based input representations, which produce varying image dimensions depending on environment size. RLoc's results show significant degradation across different environments increase the errors by more than $3\times$\cite{zhang2024rloc} owing to `bandwidth mismatch' and `center frequency mismatch'. To demonstrate this capability, we train on multiple environments and evaluate on either completely unseen environments or held-out regions within the training environments.

Figure~\ref{fig:generalization} presents eight experimental configurations. The first six involve training on five environments and testing on an unseen sixth environment. The final two experiments evaluate training exclusively on 40 MHz data (tested on held-out 40 MHz data) and training on all datasets combined (tested on unseen portions of all six environments).

Across all cross-environment experiments, median localization errors remain below 210 cm, with 50\% of the experiments achieving sub-meter median error. Additionally, the 90th percentile error across all configurations remains below 370 cm, with the best-performing configurations achieving errors as low as 155 cm, further demonstrating robust generalization across diverse environmental conditions. While 210 cm may be insufficient for precise localization of small devices, it is well-suited for indoor navigation applications. Particularly noteworthy is \name's performance when trained on environments 1, 2, 3, 5, and 6 and tested on environment 4: the system achieves 85 cm median error and maintains 90th percentile error below 210 cm. This demonstrates the network's ability to learn robust representations across both 40 MHz and 80 MHz bandwidths and varying environment sizes while maintaining sub-meter accuracy.\looseness-5

The final two experimental configurations (3-6|3-6 and 1-6|1-6) further validate \name's scalability. When trained exclusively on 40 MHz datasets (3-6|3-6), the system achieves 53 cm median error with a 90th percentile error under 200 cm. Similarly, training on all six datasets and testing on unseen portions (1-6|1-6) of these datasets yields 58 cm median error and a $90^{th}$ percentile error of 150 cm. These results demonstrate that with sufficient diverse training data, \name can achieve very high localization accuracy across bandwidths, scenario layouts, and environment sizes. \looseness-5

\subsection{Ablation Studies}\label{sec:ablation}
We now aim to motivate the purpose of each of our components in our system. All the following tests are trained and tested on \textbf{scenario E1} with 80\% of the data used for training, and 20\% for testing. The dataset consists of $\sim$20000 data points consisting of 80 MHz CSI data from 4 APs, with 1 AP behind a wall in 1500 sq. ft. indoor environment.

\subsubsection{Evidence for Federated Learning} 
In Fig.~\ref{fig:ablation}a, we compare federated and non-federated learning results. Both models use the same encoder, decoder, and geometric loss, differing only in training approach: the federated model follows FedAvg \cite{fedavg}, while the non-federated uses standard deep learning training. The federated model does show slight improvement over a non-federated model which stems from the regularization effect of federated learning. By averaging the encoder weights every 10 batches, this prevents overfitting and enhances generalization to unseen data. However, the non-federated model fine-tunes a separate encoder for each AP, achieving comparable accuracy but introducing a privacy vulnerability: an attacker could potentially identify which AP generated a given encoder output based on AP-specific encoder characteristics.

\subsubsection{Evidence for Huber Weights}
Figure~\ref{fig:ablation}b evaluates three loss functions—L1 Loss, MSE Loss, and Huber Loss—within the federated \name architecture. The CDF curves show that Huber Loss closely matches MSE Loss performance, with both significantly outperforming L1 Loss. Specifically, Huber Loss reduces median localization error by approximately 15 cm compared to L1 Loss, showcasing the benefits that motivated its selection.

\subsubsection{Evidence for Geometric Loss} 
Figure~\ref{fig:ablation}c demonstrates that our geometric loss significantly outperforms direct-AoA loss, reducing median localization error by approximately 23 cm. This advantage becomes more pronounced at higher error percentiles, with geometric loss achieving 45 cm lower error at the 90th percentile. The geometric loss enforces geometric consistency by penalizing angle predictions that are incompatible with the estimated location across all APs. In contrast, direct-AoA loss optimizes each angle independently, making it vulnerable to large localization errors when even a single AP produces a poor angle estimate.






\section{Related Works}
WiFi-based approaches are particularly attractive because they leverage existing infrastructure without requiring specialized hardware. The field has evolved to address complex challenges including multipath propagation, signal attenuation, and dynamic environmental factors. However a crucial gap remains: privacy and generalizability.

In \name, we provide machine learning and federated learning based solutions for generalizable, robust, and private indoor location estimates. So, there is existing literature from wireless localization, wireless privacy, and federated learning that we need to discuss as part of the related works.

\subsection{Wireless Localization}

Early indoor localization systems used mainly received signal strength indicator (RSSI) values to estimate the position of the user based on signal attenuation between access points (APs) \cite{ez,IEEE80211}. Although easy to implement due to RSSI availability on commonplace hardware, their accuracy tended to be limited by environmental factors such as multipath, interference, and dynamic obstructions. Thus, we see that median localization errors often exceeded 2--5 meters in most practical settings \cite {rss3, ez, rss4}. To improve accuracy, fingerprinting-based approaches emerged as a viable alternative~\cite{horus, finger2}. Although some fingerprinting systems have demonstrated sub-meter median errors in controlled settings \cite{deepfi}, they tend to suffer from high deployment costs. Recalibration is often required whenever the environment changes, such as with movement of furniture or crowd density, making them unsuitable for large-scale or dynamic environments.

These limitations in signal-strength modeling and fingerprint maintenance motivated a shift toward learning-based methods that can infer spatial structure from raw wireless measurements, leading to the use of CSI and deep neural networks in modern indoor localization systems. 
Channel state information (CSI) has been seen as a key signal representation for improving indoor localization using a commodity WiFi infrastructure \cite{csi_survey, deepwifi_csi_net, kotaru2015spotfi,chronos,soltanaghaei2018multipath,arraytrack,xie2018md,xie2016xd,xie2018swan,ubicarse,pizarro2021accurate,nandakumar20183d}. So unlike the coarse signal strength indicators with RSSI, CSI is able to provide fine-grained phase and amplitude information across subcarriers and antennas. Previous work has demonstrated sub-meter localization using CSI by applying signal processing techniques or machine learning models \cite{yang2023, tofnet, Chen2019WiFiLSTM, Gong2018RoArray, Gong2018SiFi,dloc,zhang2024rloc}.

DLoc \cite{dloc} employs an image-to-image translation approach for localization, utilizing AoA–ToF heatmaps derived from CSI data as inputs to a convolutional neural network.  However, DLoc generates location estimates through dense regression over the entire spatial map, predicting probability distributions across all grid locations without relying on explicit geometric constraints. Consequently, while DLoc is effective within a single environment, its predictions lack geometric interpretability and generalization properties. RLoc attempts at tackling the generalization problem by tackling the non-linear AoA error mapping but do not generalize well across different bandwidths or infrastructure deployments. \name overcomes all of these problems through its geometric loss based learning and 

Recent learning-based approaches can estimate AoA directly from CSI without hardware synchronization and implicitly compensate for hardware imperfections \cite{Liu2018AoADNN, Papageorgiou2021DOANoisy}. However, direct angle regression suffers from a fundamental optimization problem: at extreme angles ($\pm 90^\circ$), the gradient of standard losses vanishes due to directional ambiguity in antenna geometry, causing slow convergence or biased predictions.
Some learning-based localization systems avoid explicit ray modeling by using deep networks to map input features to location heatmaps\cite{wang}. While flexible, these approaches lack interpretability and do not enforce physical constraints relating predicted AoAs to actual geometry, limiting robustness in multipath environments.

\name combines neural networks for robust AoA prediction with geometric constraints through a differentiable ray intersection solver, addressing both the generalization limitations of dense regression and the vanishing gradient problem in angle regression. This hybrid approach strengthens cross-environment generalization and multipath robustness, and we achieve privacy by performing computation in a decentralized way rather than on a central server.

\subsection{Wireless Privacy}

Existing work on wireless localization privacy largely focuses on obfuscating the physical-layer signals that attackers rely on. One line of approaches introduces additional radios to distort or relay signals, as in Aegis~\cite{yao2018aegis}, PhyCloak~\cite{qiao2016phycloak}, and WiAdv~\cite{zhou2022wiadv}, which interfere with amplitude, phase, or learned signal features to confuse localization systems. Another direction uses smart surfaces or environmental modifications to alter propagation paths: electromagnetic shielding or jamming can disrupt RSS-based localization~\cite{zhu2018tu, sun2022sok}, while systems like RF-Protect~\cite{shenoy2022rf} and IRShield~\cite{staat2022irshield} use mmWave reflectors or Wi-Fi–band intelligent surfaces to inject phase noise, though these solutions often require significant deployment effort or degrade communication quality. Beyond physical-layer manipulation, work analyzing the 802.11 Fine Timing Measurement (FTM) protocol~\cite{ibrahim2018verification, yu2022precise} highlights new privacy risks, with attacks such as WiPeep~\cite{wipeep} exploiting protocol loopholes and defenses advocating formal Wi-Fi protocol hardening~\cite{schepers2022privacy, ftm_secure}. Additional efforts target machine-learning-based sensing vulnerabilities (e.g., RAFA~\cite{liu2023exploring}) or higher-layer protections like MAC address randomization~\cite{zhong2004privacy, alaggan2018privacy} and hardware fingerprint obfuscation~\cite{givehchian2023practical}, though these approaches are often limited to specific technologies or sacrifice utility. In contrast, \name tackles the fundamental problem of bringing the control of locating the user location on the user device instead of on the routers or the third-part servers that most of the above systems try to tackle.

\subsection{Federated Learning for Wireless Sensing}

Existing CSI-based localization methods typically require centralized data collection and training. However, such centralized approaches can be a detriment when it comes to privacy concerns and hence hinders scalability across users and devices. Federated learning (FL) has emerged as a privacy-centered paradigm by enabling decentralized training across multiple clients without sharing raw data~\cite{fedavg, bonawitz2019federatedlearningscaledesign, Li_2020}.
Early work applied federated learning to WiFi localization using RSSI fingerprints, allowing  for privacy-preserving crowdsourced training ~\cite{9148111, yin2020fedlocfederatedlearningframework}. However, these approaches inherited RSSI's accuracy limitations. 

The more recent works have extended federated learning to CSI-based systems, FedPos~\cite{10005038} addresses cross-environment generalization by aggregating only encoder parameters while maintaining personalized classifiers, achieving around 42 cm median error with cross-environment learning across only 64 distinct user location, making the dataset too sparse and customized for a more wide-scale deployment. 

Centralized location inference creates vulnerabilities: server breaches and 
operator access expose user locations~\cite{10.1145/3369812}. Differential 
privacy mitigates this through noise injection~\cite{9060228}, but stronger 
privacy requires larger noise, introducing accuracy-utility tradeoffs.

Unlike prior federated and differential privacy systems, \name decentralizes both training and inference.  During training, encoders at multiple APs aggregate parameters using federated 
averaging across multiple instances of user device measurements, enabling robust feature learning across environments. Critically, location 
computation occurs on user devices rather than centralized servers, eliminating 
infrastructure-level vulnerabilities and removing the accuracy-utility tradeoffs 
inherent in differential privacy approaches.
\section{Limitations and Future Work}\label{sec:conlusion}
In this article, we have presented a privacy-driven approach to indoor Wi-Fi-based localization that outperforms the state-of-the-art models and is able to generalize across environments and bandwidths. \name achieves these goals by utilizing a federated learning approach in conjunction with a geometric-inspired loss function to obfuscate user location data while having the ability to generalize across scenarios. The current implementation serves as a baseline framework for privacy-driven IoT sensing systems. However, such systems do have their limitations and future work is needed to provide more-accurate and secure solutions that protect users in busy spaces. We conclude with a discussion of the limitations of our work and ideas for future work. 

\noindent\textbf{Data Constraints --}
\name has been evaluated in a diverse set of indoor scenarios and Wi-Fi router configurations. However, there is no open-sourced large-scale environment dataset that can replicate the scale of a true office building, hospital, or industrial areas. We recognize that these spaces may present unique challenges that we may not be aware of and we hope to perform large-scale data collection to further strengthen our work for real-world use. We also want to highlight that we only have 80 Mhz and 40 Mhz data available, but modern Wi-Fi standards allow for bandwidths up to 320 Mhz which may present unique opportunities for further localization improvement. 

\noindent\textbf{Compromised User Device Attacks --}
While \name is able to contain the user location data solely on their device, we cannot prevent location data being taken if the user device is compromised. This is a fundamental privacy challenge that requires more work to ensure the location stays completely obfuscated on the user device memory until it is used. 
\bibliographystyle{ACM-Reference-Format}
\bibliography{reference}

\end{document}